\documentclass[a4paper]{article}

\usepackage{INTERSPEECH2022}
\usepackage{hyperref}
\usepackage{amssymb}
\usepackage{multirow}
\usepackage{booktabs}

\hypersetup{
	colorlinks=true,
	linkcolor=black,
	citecolor=black,
	urlcolor=black,
}

\title{LightHuBERT: Lightweight and Configurable Speech Representation Learning with Once-for-All Hidden-Unit BERT}
\name{Rui Wang$^{1,\dag}$, Qibing Bai$^{2,\dag}$, Junyi Ao$^3$, Long Zhou$^{4}$, Zhixiang Xiong$^{1}$,\\Zhihua Wei$^{1}$, Yu Zhang$^{2,5}$, Tom Ko$^{6}$, Haizhou Li$^{3}$\thanks{$^{\dag}$Equal contribution. Corresponding author: Zhihua Wei (zhihua\_wei@tongji.edu.cn).}}

\address{
$^1$Department of Computer Science and Technology, Tongji University\\
$^2$Department of Computer Science and Engineering, Southern University of Science and Technology\\
$^3$School of Data Science, The Chinese University of Hong Kong (Shenzhen) \\
$^4$Microsoft~$^5$Peng Cheng Laboratory~$^6$ByteDance AI Lab}
\email{}
\begin{document}

\maketitle
\begin{abstract}
Self-supervised speech representation learning has shown promising results in various speech processing tasks. 
However, the pre-trained models, e.g., HuBERT, are storage-intensive Transformers, limiting their scope of applications under low-resource settings.
To this end, we propose LightHuBERT, a once-for-all Transformer compression framework, to find the desired architectures automatically by pruning structured parameters.
More precisely, we create a Transformer-based supernet that is nested with thousands of weight-sharing subnets and design a two-stage distillation strategy to leverage the contextualized latent representations from HuBERT.
Experiments on automatic speech recognition (ASR) and the SUPERB benchmark show the proposed LightHuBERT enables over $10^9$ architectures concerning the embedding dimension, attention dimension, head number, feed-forward network ratio, and network depth.
LightHuBERT outperforms the original HuBERT on ASR and five SUPERB tasks with the HuBERT size, achieves comparable performance to the teacher model in most tasks with a reduction of 29\% parameters, and obtains a $3.5\times$ compression ratio in three SUPERB tasks, e.g., automatic speaker verification, keyword spotting, and intent classification, with a slight accuracy loss. The code and pre-trained models are available at {\url{https://github.com/mechanicalsea/lighthubert}}.
\end{abstract}

\section{Introduction}

Self-supervised speech representation learning has shown that pre-trained models benefit from abundantly available unannotated data and produce promising results on various speech processing tasks \cite{ao2021speecht5,chung2018speech2vec,chuang2019speechbert,song2019speech,NEURIPS2020_92d1e1eb,wang2021unispeech,chung2021w2v}. 
Most speech applications require models to interact with humans or machines, therefore demanding real-time performance for a better user experience. 
However, many real-world devices, e.g., smartwatches, mobile phones, tablets, audio-visual robots, and industrial PCs, are highly constrained by memory and battery. 
This prevents the pre-trained speech models from actual deployments.

Building a compression framework for pre-trained models to meet various resource constraints can be essential for the development of speech pre-training.
It can utilize pre-trained models trained on large-scale unlabeled data to boost the performance of compressed networks in downstream tasks and lower entry barriers for pre-training related works.
However, two challenges remain when compressing pre-trained models without compromising their effectiveness:
(1) a lightweight or sparse network often suffers from a significant performance drop and demands a high computational cost due to multiple rounds of pre-training;
(2) existing compression methods on pre-trained speech models such as DistilHuBERT \cite{chang2022distilhubert} are deficient under various resource constraints. 
Designing a general-purpose compression framework that enables different sizes of models under a tolerable training time is not well explored.

Inspired by the once-for-all approach \cite{Cai2020}, we propose LightHuBERT, an efficient model compression framework for lightweight and configurable speech pre-training, which consists of a once-for-all Transformer, a contextualized latent representation distillation objective, and a two-stage training strategy. 
Specifically, similar to \cite{Chen2021}, we build a weight-sharing Transformer supernet with channel-variable convolutional positional embeddings that can enable a Transformer architecture to scale in the embedding dimension, the head number, the feed-forward network (FFN) ratio, and the network depth. 
We introduce a pre-training distillation loss and perform masked self-supervised learning, where the student predicts contextualized representations within masked time steps that contain context information from a pre-trained teacher, the HuBERT \textsc{Base} model \cite{Hsu2021}.
A two-stage training strategy is designed to further improve the performance.
Experiments on the automatic speech recognition and the SUPERB benchmark show that LightHuBERT attains superior results under various parameters.

The contributions of this paper are summarized as follows.

\begin{itemize}
    \item We propose a lightweight and configurable model compression framework for speech pre-training to address the challenge of applying pre-trained models to various computational resources.
    \item 
    We propose (1) a once-for-all Transformer with two individual supernets, (2) the contextualized latent representations to transfer knowledge, and (3) a two-stage training strategy to improve the performance.
    \item We demonstrate the effectiveness of the proposed method with pre-trained HuBERT \textsc{Base} on automatic speech recognition task and the SUPERB benchmark.
\end{itemize}

\begin{figure*}
    \centering
    \includegraphics[width=16.2cm]{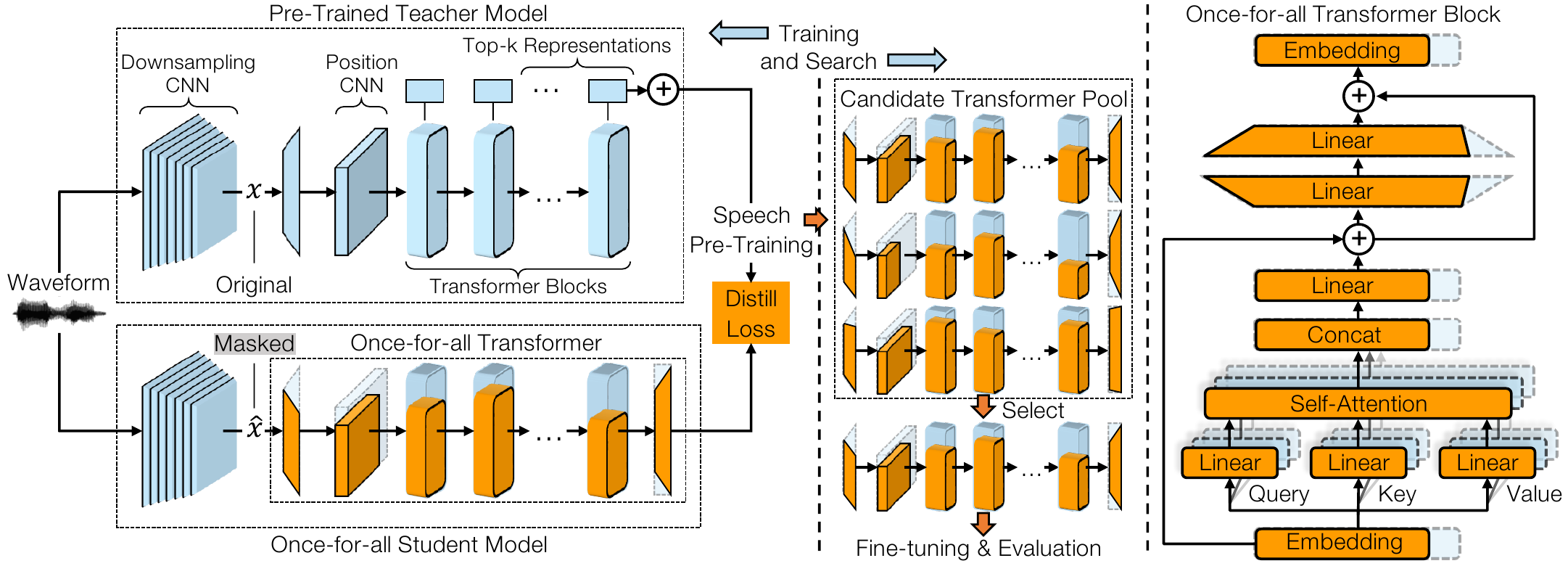}
    \caption{Model compression for speech pre-training. All orange components are dynamic and share weights with large ones. Since this work focuses on pruning the encoder in speech pre-training, we keep the downsampling convolutional network static. \textbf{Left:} Given a pre-trained speech model, we distill a once-for-all Transformer containing all possible dimensions through weight sharing. We randomly sample different pairs of these variable dimensions at each training step so that one Transformer can support all architectures. \textbf{Middle:} We then select many sub-Transformers with different networks from the once-for-all Transformer, validate them without retraining, and finally select the best architecture given the parameter constraint to perform an evaluation. \textbf{Right:} The once-for-all Transformer block supports multiple dynamic dimensions: embedding dimension, attention dimension, head number, FFN ratio, and network depth.}
    \label{fig:framework}
\end{figure*}

\section{Related Work}


Large-scale pre-trained models such as wav2vec 2.0 \cite{NEURIPS2020_92d1e1eb}, HuBERT \cite{Hsu2021}, WavLM \cite{chen2021wavlm}, SpeechT5 \cite{ao2021speecht5}, and data2vec \cite{baevski2022data2vec} have drawn much attention in the speech communities, due to their excellent generalization capability across various speech applications and efficient use of large-scale unlabeled data. SUPERB \cite{Yang2021} has appeared as a popular benchmark to evaluate the performance of pre-trained models, where the pre-trained encoder is frozen and shared to investigate the capability of representations on various speech tasks, including phoneme recognition (PR), automatic speech recognition (ASR), keyword spotting (KS), query by example spoken term detection (QbE), speaker identification (SID), automatic speaker verification (ASV), speaker diarization (SD), intent classification (IC), slot filling (SF), and emotion recognition (ER). 


Reducing architectural complexity of large-scale pre-trained models has become an indispensable research endeavor \cite{sanh2019,chang2022distilhubert,Aguilar2020,Lai2021,zhang2021you,yu2021unified,Zafrir2021,peng2021shrinking}. 
DistilHuBERT \cite{chang2022distilhubert} is proposed to distill hidden representations from HuBERT \textsc{Base}. 
It creates a two-layer Transformer while retaining most performance on ten SUPERB tasks. 
Unlike the DistilHuBERT model, which creates a single compressed network, we distill a once-for-all network to deploy under various computational resources. 
PARP \cite{Lai2021} is a magnitude-based unstructured pruning method that removes insignificant weights of wav2vec 2.0 based on a sparse network on low-resource ASR with the monolingual and cross-lingual transfer.
Unlike PARP, we prune structured groups of weights, therefore, avoiding irregular sparse matrix operations that are hard to accelerate on hardware \cite{gale2019state}.

Once-for-all approaches \cite{Cai2020} aim to create a weight-sharing supernet and obtain a huge number of architectures for efficient deployment with different resource constraints while maintaining the same level of accuracy as those trained independently \cite{deng2020model,he2021automl,kuutti2020survey,lu2021neural}. Examples can be found in image processing \cite{Cai2020,Li2020}, natural language processing \cite{NEURIPS2020_6f5216f8}, and speech processing \cite{wang2021efficienttdnn}. 
Similarly, AutoFormer \cite{Chen2021} utilizes weight entanglement to supernet training and enables Transformer blocks to share weights for their common in each layer. 
It allows a large number of subnets in the supernet to be as well-trained as ones trained from scratch.
Unlike AutoFormer, we introduce a two-stage distillation to improve the supernet while utilizing a task-agnostic objective to learn speech representations.

\section{LightHuBERT}

As shown in Figure \ref{fig:framework}, in this section, we propose LightHuBERT, a task-agnostic compression framework for reducing the model size of the Transformer encoder in speech pre-training. Specifically, we propose a once-for-all Transformer that enables automatic architecture search by pruning structured groups of weights. We transfer contextualized latent representations to sub-Transformers. A two-stage training strategy is proposed to improve the compressed models.

\subsection{Once-for-All Transformer}

Once-for-all Transformer refers to a Transformer architecture that contains various sub-Transformers, where different architectures share weights in a scaling manner. For example, a 256-dim linear layer is nested in a 512-dim linear layer. We design a once-for-all Transformer that contains five variable dimensions similar to AutoFormer \cite{Chen2021}: embedding dimension, attention dimension, head number, FFN ratio, and network depth. We constrain the attention dimension (i.e., key, query, and value matrices) as the 64$\times$ head number.

Since the interference between small and large networks degrades the performance of the large ones \cite{wang2021efficienttdnn}, we create two supernets to investigate models with significantly different model sizes, as shown in Table \ref{tab:ofa_transformer}. Two supernets retain most Transformer blocks because existing Transformer-based speech pre-training usually applies deep networks to learn from large-scale unlabeled data and maintain superior performance, such as 12-layer HuBERT \textsc{Base} and 24-layer HuBERT \textsc{Large}. To clarify, we denote $a_\text{Largest}$, $a_\textsc{Base}$, and $a_\textsc{Small}$ as shown in Table \ref{tab:fui}.

\begin{table}[t]
 \setlength{\abovecaptionskip}{3pt}
 \setlength{\belowcaptionskip}{-0pt}
    \renewcommand\arraystretch{1.1}
    \centering
    \caption{Two supernets of once-for-all Transformers.}
    \scriptsize
    \begin{tabular}{lcc}
        \toprule
        & Small & Base \\
        \midrule
        Embedding Dim & \{256, 384, 512\} & \{512, 640, 768\} \\
        Head Number & \{4, 6, 8\} & \{8, 10, 12\} \\
        FFN Ratio & \{3.0, 3.5, 4.0\} & \{3.5, 4.0\} \\
        Network Depth & \{10, 11, 12\} & \{12\} \\
        \midrule
        Parameters Range & 11M -- 45M & 41M -- 95M \\
        Subnets Size & $9.5\times10^{11}$ & $6.5\times10^9$ \\ 
        \bottomrule
    \end{tabular}
    \label{tab:ofa_transformer}
\end{table}

\subsection{Pre-Training Distillation}

We employ a masking-based pre-training distillation to transfer the knowledge of a pre-trained model. Specifically, we mask spans of latent speech representations in the student model and make the student model predict masked parts as the output of the teacher model. Inspired by \cite{baevski2022data2vec}, we introduce contextualized representations as the training target, i.e., average top-k normalized latent representations, where we set $\text{k}=8$ as \cite{baevski2022data2vec}. Unlike self-distillation in \cite{baevski2022data2vec}, we leverage a pre-trained speech model as the teacher. Formally, given a downsampled audio sequence $x$, the student is to minimize the L1 distance within masked time steps $\mathcal{M}$ as
\begin{equation}\label{equ:objective}
    \mathcal{L}\left(f^{t}(x),f^{s}(\hat{x})\right)=\frac{1}{|\mathcal{M}|}\sum_{i\in\mathcal{M}}{\left|\bar{f_i^{t}}(x)-f_i^{s}(\hat{x})\right|},
\end{equation}
where $f^t(\cdot)$ denotes the teacher, $f^s(\cdot)$ denotes the student, $\hat{x}$ is the masked $x$ with a masking probability of $p=0.65$ as \cite{baevski2022data2vec}, and $\bar{f_i^{t}}(\cdot)$ denotes the training target at the $i$-th time step.

\subsection{Two-Stage Training}

To improve the performance of different weight-sharing architectures in the once-for-all Transformer, we propose a two-stage training strategy as follows.

\begin{itemize}
    \item \textbf{Stage 1 - Distillation.} We train the largest architecture $a_\text{Largest}$ of the once-for-all Transformer from scratch via the loss function of the pre-training distillation.
    \item \textbf{Stage 2 - Once-for-All Training.} We implement the once-for-all training on the supernet initialized by distilled weights. Specifically, we randomly sample a subnet from the supernet at each forwarding propagation during the supernet training.
\end{itemize}

The trained weights derived from Stage 1 serve as the initialization of Stage 2. Compared to existing pre-training objectives that force the top-layer representations to fit the targets, the pre-training distillation utilizes contextualized representations that provide receptive fields in different ranges and feature aggregation with various resolutions, which can be helpful to train subnets. For clarity, we define OFA HuBERT and LightHuBERT as shown in Table \ref{tab:fui}.

\begin{table}[t]
    \renewcommand\arraystretch{1.1}
    \centering
    \caption{Frequently used items.}
    \resizebox{\linewidth}{!}{
    \begin{tabular}{ll}
        \toprule
        Symbol & Description \\
        \midrule
        \large $a_\text{Largest}$ & The largest architecture as the HuBERT $\textsc{Base}$ size. \\
        \large $a_\textsc{Base}$ & 12-layer 640-embedding 10-head 2560-FFN subnet. \\
        \large $a_\textsc{Small}$ & 12-layer 384-embedding 6-head 1536-FFN subnet. \\
        OFA HuBERT & Once-for-all training initialized with the pre-trained HuBERT. \\
        LightHuBERT & Two-stage training for a once-for-all HuBERT. \\
        \bottomrule
    \end{tabular}}
    \label{tab:fui}
\end{table}

\section{Experiments}

We conduct our method with pre-trained HuBERT \textsc{Base}. Pre-training models and the 10 hours ASR are conducted in fairseq \cite{ott2019fairseq}. The SUPERB tasks are implemented with S3PRL \cite{Yang2021}.

\subsection{Experimental Setup}

\noindent\textbf{Model.} The HuBERT \textsc{Base} model has a 7-layer temporal convolution, 1-layer convolutional position, and a 12-layer Transformer encoder. The pre-trained weights are downloaded from \cite{ott2019fairseq}. Distilled student models have a similar architecture with an additional prediction head that predicts the training targets.

\noindent\textbf{Datasets.} For the pre-training task, we use a total of 960 hours of LibriSpeech audio \cite{Panayotov2015}. For the ASR task, we fine-tune each model on the 10 hours labeled split of Librilight \cite{Kahn2020} and report the word error rate (WER) without a language model. For the SUPERB benchmark \cite{Yang2021}, we evaluate the models on ten tasks with the officially provided datasets, training recipes, and evaluation protocols, including PR, ASR, KS, QbE, SID, ASV, SD, IC, SF, and ER.

\noindent\textbf{Pre-Training.} The default pre-training is initialized with either the publicly released or the re-implemented weights. We pre-train once-for-all Transformers on 8 V100 GPUs with a batch size of around 119 seconds of audios per GPU for 200k steps, where the downsampling CNN is frozen to significantly reduce training time. Distilling $a_\text{Largest}$ from scratch is implemented on 32 GPUs as the HuBERT \textsc{Base} training recipe.

\noindent\textbf{Search.} We randomly search for subnets of the pre-trained models with the pre-trained distilled objective given parameters. For each pre-trained model, we search for 1,000 subnets. Besides, the minimal and maximal subnets are evaluated to estimate the potential performance bounds. Three architectures, i.e., $a_\text{Largest}$, $a_\textsc{Base}$, and $a_\textsc{Small}$, are selected manually for evaluation on the SUPERB benchmark.

\begin{table}[t]
    \centering
    \caption{ASR results on $a_\text{Largest}$ between once-for-all Transformers and HuBERT. $^\star$HuBERT \textsc{Base} is reproduced.}
    \label{tab:largest_asr}
\resizebox{\linewidth}{!}{
    \begin{tabular}{lcccc}
        \toprule
        Method & dev-clean & dev-other & test-clean & test-other \\
        \midrule
        $^\star$HuBERT \textsc{Base} \cite{Hsu2021} & 9.6 & 16.3 & 9.6 & 16.9 \\ 
        OFA HuBERT & 10.1 & 17.4 & 10.3 & 17.9 \\ 
        LightHuBERT & 9.0 & 16.4 & 9.3 & 16.6 \\ 
        LightHuBERT $a_\textsc{Base} $ & 9.4 & 17.3 & 9.6 & 17.5 \\ 
        LightHuBERT Stage 1 & \textbf{8.6} & \textbf{14.5} & \textbf{8.7} & \textbf{14.8} \\ 
        \bottomrule
    \end{tabular}
}
\end{table}

\begin{table*}[ht]
 \setlength{\abovecaptionskip}{3pt}
 \setlength{\belowcaptionskip}{-0pt}
 \renewcommand\arraystretch{1.0}
 \centering
 \caption{Speech pre-training evaluation on the SUPERB benchmark. Metrics include accuracy (Acc\%), phoneme error rate (PER\%), word error rate (WER\%), maximum term weighted value (MTWV), F1 score (F1\%), concept error rate (CER\%), equal error rate (EER\%), and diarization error rate (DER\%). ``Overall Score" denotes the average scores of all tasks as \cite{chen2021wavlm}. ``Paral." denotes paralinguistics. ``Distill." denotes distillation. $^\star$HuBERT Teacher is reproduced and is slightly different from the released HuBERT.}
 \label{tab:superb}
\resizebox{\linewidth}{!}{
 \begin{tabular}{lcccccccccccccc}
 \toprule
 \multirow{3}*[-2pt]{Method} & \multirow{3}*[-2pt]{Params} & \multirow{3}*[-2pt]{\begin{tabular}{@{}c@{}}Overall \\ Score\end{tabular}} & \multicolumn{5}{c}{Content} & \multicolumn{3}{c}{Speaker} & \multicolumn{3}{c}{Semantics} & Paral. \\
 \cmidrule(r){4-8}\cmidrule(r){9-11}\cmidrule(r){12-14}\cmidrule(r){15-15}
 & & & PR & \multicolumn{2}{c}{ASR (WER)} & KS & QbE & SID & ASV & SD & IC & \multicolumn{2}{c}{SF} & ER \\
 \cmidrule(r){4-4}\cmidrule(r){5-6}\cmidrule(r){7-7}\cmidrule(r){8-8}\cmidrule(r){9-9}\cmidrule(r){10-10}\cmidrule(r){11-11}\cmidrule(r){12-12}\cmidrule(r){13-14}\cmidrule(r){15-15}
 & & & PER $\downarrow$ & w/o $\downarrow$ & w/ LM $\downarrow$ & Acc $\uparrow$ & MTWV $\uparrow$ & Acc $\uparrow$ & EER $\downarrow$ & DER $\downarrow$ & Acc $\uparrow$ &  F1 $\uparrow$ & CER $\downarrow$ & Acc $\uparrow$\\
 \midrule
 HuBERT \textsc{Base} \cite{Hsu2021} & 95M & 80.8 & 5.41 & 6.42 & 4.79 & 96.30 & 0.0736 & 81.42 & 5.11 & 5.88 & 98.34 & 88.53 & 25.20 & 64.92 \\
 DistilHuBERT \cite{chang2022distilhubert} & 23M & 75.9 & 16.27 & 13.34 & 9.21 & 95.98 & 0.0511 & 73.54 & 8.55 & 6.19 & 94.99 & 82.57 & 35.59 & 63.02 \\
 \midrule
 $^\star$HuBERT Teacher & 95M & 81.1 & 5.23 & 6.62 & 4.96 & 96.66 & \textbf{0.0879} & \textbf{82.90} & \textbf{4.94} & \textbf{5.45} & 98.21 & \textbf{88.81} & \textbf{25.30} & 64.68 \\ 
 Stage 1 Distill. {\large $a_\text{Largest}$} & 95M & 81.0 & \textbf{4.15} & \textbf{5.71} & \textbf{4.20} & \textbf{96.82} & 0.0737 & 80.01 & 5.14 & 5.51 & \textbf{98.50} & 88.44 & 25.92 & \textbf{66.25} \\
 LightHuBERT {\large $a_\text{Largest}$} & 95M & 80.4 & 4.56 & 6.43 & 4.80 & 96.40 & 0.0642 & 77.58 & 5.43 & 5.85 & 98.21 & 88.78 & 25.32 & 64.93 \\
 LightHuBERT {\large $a_\text{Base}$} & 68M & 80.4 & 4.71 & 6.72 & 4.97 & 95.75 & 0.0767 & 77.24 & 5.55 & 5.73 & 98.00 & 88.79 & 26.06 & 65.55 \\ 
 LightHuBERT {\large $a_\text{Small}$} & 27M & 79.1 & 6.60 & 8.33 & 6.04 & 96.07 & 0.0764 & 69.70 & 5.42 & 5.85 & 98.23 & 87.58 & 26.90 & 64.12 \\ 
 \bottomrule
 \end{tabular}
}
\vspace{-8pt}\end{table*}

\subsection{Automatic Speech Recognition}

We conduct the once-for-all Transformer with HuBERT \textsc{Base} on the ASR task and evaluate the largest architecture $a_\text{Largest}$ from the base supernet. As shown in Table \ref{tab:largest_asr}, the model of LightHuBERT at stage 1 outperforms the HuBERT teacher, which indicates the effectiveness of our distillation targets.
LightHuBERT achieves superior performance than the OFA HuBERT, and its $a_\textsc{Base}$ achieves comparable performance to the HuBERT teacher with 29\% reduced parameters, which suggests the two-stage strategy helps learn subnets.
On the other hand, the performance of $a_\text{Largest}$ slightly degrades in both OFA HuBERT and LightHuBERT after the once-for-all training, probably due to the interference between small and large networks caused by the once-for-all training.

\begin{figure}[t]
    \centering
    \setlength{\abovecaptionskip}{3pt}
    \setlength{\belowcaptionskip}{-0pt}
    \includegraphics[width=7.2cm]{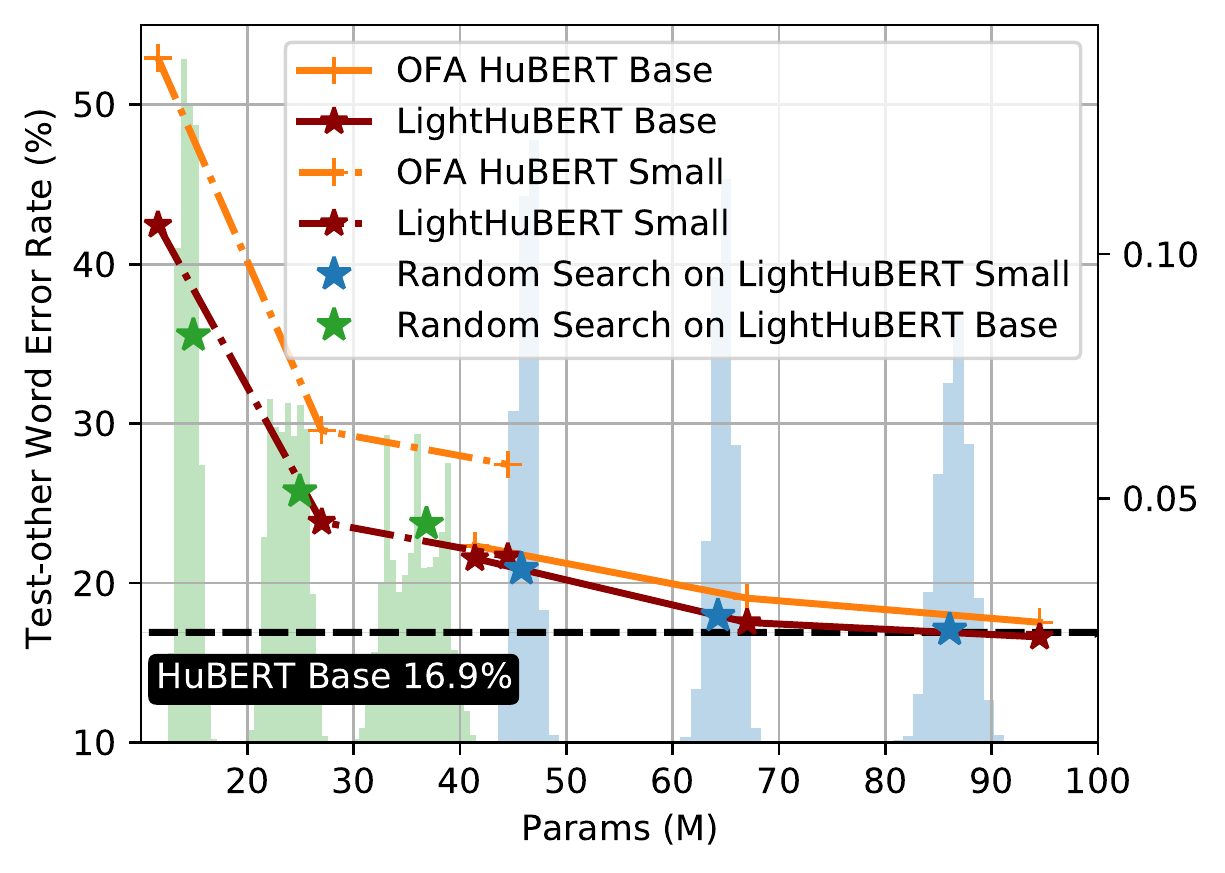}
    \caption{ASR results of the test-other between LightHuBERT and OFA HuBERT. Green and blue histograms denote the distributions of parameters of subnets sampled from the small and base supernets, respectively. We evaluate six subnets found by the random search in the LightHuBERT. Three of them are found given 15M, 25M, and 37M parameters in the small supernet, and the others are found given 47M, 65M, and 87M parameters in the base supernet.}
    \label{fig:ofa_3}
\vspace{-6pt}\end{figure}

To estimate the performance of the trained supernets, we evaluate some subnets from four trained once-for-all Transformers. These pre-trained Transformers include the OFA HuBERT and the LightHuBERT from small and base supernets. These subnets include the minimal, the maximal, two manually selected, and several randomly found architectures.
The results are illustrated in Figure \ref{fig:ofa_3}. The results of different selected architectures illustrate that these once-for-all Transformers obtain many well-trained sub-architectures. All chosen architectures from LightHuBERT outperform the networks from OFA HuBERT, which represents the superiority of the two-stage training strategy.

\subsection{Universal Representation Evaluation}

To provide a comprehensive testbed for the generalizability of our compressed pre-trained models, we evaluate four models derived from LightHuBERT on the SUPERB benchmark. As shown in Table \ref{tab:superb}, we can draw the following conclusions: (1) The proposed LightHuBERT can create compressed models that retain comparable performance on the SUPERB tasks. The $a_\textsc{Small}$ with around 28\% parameter achieves only a 2 point drop in overall score, which significantly outperforms DistilHuBERT by an absolute improvement of 3.2 point. 
(2) The LightHuBERT at Stage 1 achieves superior performance than the HuBERT teacher in five tasks while maintaining comparable performance in other tasks, which demonstrates the effectiveness of the proposed pre-training distillation to learn universal speech representations.
(3) We find the $a_\textsc{Small}$ achieves $3.5\times$ compression ratio in KS, ASV, and IC tasks with a slight accuracy loss, and the $a_\textsc{Base}$ attains comparable performance with the HuBERT teacher by saving 29\% parameters in most tasks such as PR and ASR, which indicates that the model size could vary across SUPERB tasks.

\subsection{Ablation Study}

As shown in Table \ref{tab:ablation}, an ablation study is conducted to investigate the pre-training distillation and the once-for-all training by evaluating the $a_\textsc{Base}$ on the ASR task. 
The performance of the pre-trained $a_\textsc{Base}$ drops consistently without either pre-training distillation or once-for-all training, which suggests that both pre-training distillation and once-for-all contribute to improving the supernet. Both stages significantly enhance the performance of the subnet $a_\textsc{Base}$ by over 20\% relative improvement in WER, which reveals that initializing weights and sampling diverse subnets during training help obtain well-trained networks.

\begin{table}
 \setlength{\abovecaptionskip}{3pt}
 \setlength{\belowcaptionskip}{-0pt}
 \centering
 
 \caption{Ablation study with $a_\textsc{Base}$ from the base supernet. We report WER (\%) on the ASR task without a language model.}
 \label{tab:ablation}
\resizebox{\linewidth}{!}{
 \begin{tabular}{lrr}
 \toprule
 Method & test-clean $\downarrow$ & test-other $\downarrow$  \\
 \midrule
 LightHuBERT & 9.6 & 17.5 \\ 
 \quad - Stage 1 Distillation & 13.3 (+3.7) & 22.4 (+4.9) \\ 
 \quad - Stage 2 Once-for-All Training & 12.2 (+2.5) & 22.8 (+5.3) \\ 
 \bottomrule
 \end{tabular}
}
\vspace{-8pt}\end{table}

\section{Conclusion}

In this paper, we propose LightHuBERT, a once-for-all Transformer compression framework to produce many pre-trained models in different sizes, making it available to run pre-trained models under various computational resources.
Experiments with the pre-trained HuBERT \textsc{Base} on the 10 hour ASR task and the SUPERB benchmark demonstrate the effectiveness of the proposed LightHuBERT. 
Considering different parameters, the distilled model achieves superior performance compared to the HuBERT \textsc{Base} teacher in most speech tasks with the size of the teacher model. 
LightHuBERT obtains comparable performance to the HuBERT teacher in most tasks with 29\% reduced parameters. 
We achieve a $3.5\times$ compression ratio in ASV, KS, and IC tasks with a slight accuracy loss while outperforming DistilHuBERT by an absolute improvement of 3.2 point in terms of the overall score. 
In future work, we will jointly employ parameter compression \cite{wang2021exploring} and task-dependent model compression \cite{zhang2021you} to further compress the pre-trained model.

\section{Acknowledgements}
\label{sec:acknowledgements}

The work is partially supported by the National Nature Science Foundation of China (No. 61976160, 61906137, 61976158, 62076184, 62076182) and Shanghai Science and Technology Plan Project (No. 21DZ1204800) and Technology research plan project of Ministry of Public and Security (Grant No. 2020JSYJD01).

\bibliographystyle{IEEEtran}
\bibliography{mybib}

\end{document}